# Temperature dependence and mechanism of electrically detected ESR at the ν=1 filling factor of a two-dimensional electron system in GaAs Quantum Wells

Eugene Olshanetsky, Manyam Pilla, Joshua D. Caldwell, Shu-chen Liu, and Clifford R. Bowers

Chemistry Department and National High Magnetic Field Laboratory,

University of Florida, Gainesville FL 32611-7200 USA

Jerry A. Simmons, John L. Reno

Sandia National Laboratories, MS 1415, Albuquerque, NM 87185 USA

**Abstract**

Electrically detected electron spin resonance (EDESR) signals were acquired as a function of temperature in the 0.3-4.2 K temperature range in a AlGaAs/GaAs multiple quantum well sample at the $\nu = 1$ filling factor at 5.7 T. In the particular sample studied, the line width is approximately temperature independent, while the amplitude exhibits a maximum at about 2.2 K and vanishes with increased or decreased temperature. To explain the observed temperature dependence of the signal amplitude, the signal amplitude temperature dependence is calculated assuming a model based on heating. The model ascribes the resonant absorption of microwave power of the 2DES to the uniform mode of the electron spin magnetization where the elementary spin excitations at filling factor $\nu = 1$ are taken to be spin waves, while the short wavelength spin wave modes serve as a heat sink for the absorbed energy. Due to the finite thermal conductance to the surroundings, the temperature of the 2DES spin wave system is increased, resulting in a thermal activation of the longitudinal magnetoconductance. The proposed heating model correctly predicts the location of the maximum in the experimentally observed temperature dependence of the EDESR amplitude. It also correctly predicts that the signal should vanish as the temperature is increased or decreased. The results of the present study demonstrate how experimental EDESR studies can, under appropriate conditions, provide data that can be used to discriminate between competing theories for the magnetic ordering and magnetic excitations of a 2DES in the regime of the quantum Hall effect.

**Introduction**.

The physics of a two-dimensional electron system (2DES) in a magnetic field is in many respects unique due to The interplay between the electron-electron and electron Zeeman interactions in a two-dimensional electron system leads to a wide variety of interesting phenomena at sufficiently low temperatures and high applied magnetic field. Electron spin resonance (ESR) has the potential to serve as a tool for directly probing the electron spin order and dynamics that cannot be obtained by standard transport measurements that probe only the $k \to \infty$ excitations of the system. For example, the bare g-factor and its field dependence can be measured most accurately by ESR, while the spin dynamics can be directly probed via ESR measurement of the $T_1$ or $T_2$ spin relaxation times. The spin relaxation times, which are related to magnetic fluctuations, can be extracted from ESR line shape analysis, microwave power dependence, or pulsed methods. However, there are serious technical difficulties that make direct microwave absorption detection of ESR quite problematic in the case of a 2DES due to the comparatively small number of electron spins at the electron densities of interest in single or even multiple GaAs/Al$_{1-x}$Ga$_x$As quantum wells [1]. It has been previously demonstrated that high sensitivity can be obtained if the ESR absorption is detected electrically via a change induced in the magnetoresistance that can be observed under certain conditions in the regime of the quantum Hall effect [2]. This method has been successfully used in the analysis of the magnetic field dependence of the bare electron g-factor [3,4,5], the electron-nuclear interaction and nuclear-spin relaxation rates as a function of magnetic field [6] and the g-factor dependence on the top-gate voltage [7]. Although a variety of applications of EDESR have been demonstrated in quantum Hall systems, there have been no previous reports describing the underlying physics of the electrical response to ESR transitions in a 2DES. In particular, the temperature dependence of the EDESR amplitude remains to be described theoretically, even though there is at least one previous discussion in the literature [5].

The present work focuses on an experimental study of EDESR at filling factor $\nu = 1$, where the electron-electron and electron spin Zeeman interactions produce an energy splitting in the density of states of the lowest Landau level. In the simplest qualitative description of the electronic transport at $\nu = 1$, the Fermi energy is located in the middle of the energy gap due to the spin splitting, and the longitudinal conductivity, $\sigma_{xx}$, vanishes as $T \to 0$. In EDESR, a perturbation of the equilibrium electron spin polarization due to magnetic dipole driven spin-flip transitions produces a photoconductive signal. The 0.3-4.2 K temperature dependence of the signal amplitude and resonance line width of EDESR was measured in a 2D electron system in AlGaAs/GaAs quantum wells for $\nu = 1$ at 5.4-5.7 T. The temperature dependence of signal amplitude is analyzed in detail using the concept of spin waves at filling factor $\nu = 1$ since it was determined in a previous study of nuclear spin-orientation dependence of magnetoconductance that spin waves rather than Skyrmions are the elementary spin excitations in this particular sample (sample EA-124). [8]

**Samples and Experiment**

The EDESR signals were detected from both multiple and single $Al_{1-x}Ga_xAs/GaAs$ quantum well samples grown by molecular beam epitaxy. These samples have mobilities between $\mu=4.4\times10^5$-$1.2\times10^6$ cm$^2$/V·s and electron densities per layer of $N_s=(7-25)\times10^{10}$ cm$^{-2}$. Samples were patterned with a conventional Hall bar geometry with a 0.2 mm channel width and a typical voltage probe separation of 1.5 mm. The behavior of the EDESR in all of these samples was found to be qualitatively the same. The experimental data presented below was measured on sample EA-124 (21×300 A wide GaAs wells, $Al_{0.1}Ga_{0.9}As$ barriers). This particular sample has a 2D electron density per layer of $N_s=6.9\times10^{10}$ cm$^{-2}$ and a mobility of $\mu=4.4\times10^5$ cm$^2$/V·s. The sample was mounted on a rotation stage allowing the ESR condition to be obtained at a desired filling factor and magnetic field. The ESR signals were detected via a change in the longitudinal resistance $\Delta R_{xx}$ due to the spin resonance absorption of the microwaves by the 2DES. [2-10] As described previously [8,9], a double lock-in technique was used to measure $\Delta R_{xx}$. The schematic of the experimental setup is shown in fig.1. A $5\times10^{-7}$ A (RMS) AC current ($f_1$=537 Hz) is applied to the sample with a 10 MΩ resistor in series. The microwave system consists of a YIG oscillator (Micro Lambda model MLOS 1392PA) with a tunable output of 10-18 GHz connected to a doubling amplifier (DBS model DB97-0426) which produces >15dBm level (into 50Ω) at the output. The microwave components were connected together using 3.5mm coaxial connectors and adapters (Anritsu K-connectors). An absorptive PIN diode modulator (General Microwave model D1959) is used to square wave modulate the 20-36 GHz microwave field at a frequency of $f_{mod}$=11.7 Hz. The microwaves are directed to the sample via a 50 Ω, 3mm O.D semi-rigid coax. The first lock-in amplifier is set to frequency $f_1$ and measures the signal proportional to $R_{xx}$. The $R_{xx}$ signal will contain an oscillatory component of frequency $f_{mod}$ proportional to $\Delta R_{xx}$. In order to transfer this component without attenuation to the input of the second lock-in, the time constant of the first lock-in is set to 100 μs. The $\Delta R_{xx}$ signal is extracted by the second lock-in, which is set to $f_{mod}$ and a much longer time constant of 300 ms to reduce noise. The present study concerns EDESR experiments only in the vicinity of filling factor ν=1.

**Experimental results**.

A typical $\Delta R_{xx}$ vs. $B_0$ curve together with a corresponding $R_{xx}$ vs. $B_0$ trace, both of which were acquired at 1.5 K, are shown in fig. 2. For this experiment the sample was tilted at 60° with respect to the $B_0$ field in order to bring the ν=1 magnetoresistance minimum to a field of $B_0$=5.5-5.8 T. This field range corresponds to an electron spin Larmor frequency of approximately 30 GHz, a frequency that is within the range of our microwave system. A peak in $\Delta R_{xx}(B_0)$ is clearly apparent due to ESR at a field of 5.47 T. The ESR signal is superimposed on a non-resonant background contribution to $\Delta R_{xx}$. For magnetic field sweep rates of 0.448 T/min (the highest sweep rate attainable with our magnet), the EDESR signal is the same for both the up sweep and the down sweep dependences in EA-124. For slower sweep rates, however, effects due to dynamic nuclear polarization and the Overhauser shift become noticeable. In the down sweep, the EDESR peak broadens and exhibits diminished amplitude, while the signal acquired on the up sweep is narrowed. These observations are consistent with previously reported EDESR experiments involving Overhauser shifts in GaAs/AlGaAs heterostructures [6,10]. To avoid the effects of dynamic nuclear polarization and Overhauser shifts, all the spectra presented in this paper have been recorded at the maximum sweep rate of 0.448 T/min where the EDESR peak does not show any effect due to DNP. The ESR line position is given by $h\nu_0 = g\mu_B B_{0,esr}$, where $\nu_0$ is the microwave resonance frequency, $B_{0,esr}$ is the magnetic field at which the resonance is observed and $g$ is the bare g-factor for single spin flips, which in EA-124 is $g$= -0.418 at the magnetic fields of interest. The non-resonant component of $\Delta R_{xx}$ is an oscillating function of magnetic field whose main features correlate with $R_{xx}$. Figure 3 shows the typical EDESR traces measured at different temperatures. These data represent the raw $\Delta R_{xx}$ signals that include a nonresonant background contribution. In order to quantitatively analyze the temperature dependence of the signal contribution due to ESR, this nonresonant background component is removed using the following procedure. First, the portion of the raw $\Delta R_{xx}$ trace in the region of the EDESR peak was deleted. Using the eye as a guide, several sample points were inserted into the missing region to facilitate interpolation of the background signal. A standard sixth degree polynomial fit to the background signal was obtained and then subtracted from the original raw data to obtain the component due to ESR absorption. The pure EDESR trace was found to be reasonably insensitive to the particular positioning of the manually inserted sample points and is well described as a Gaussian line shape function. A standard Gaussian fit was then performed on the EDESR trace to obtain the line width and signal integral of the resonance peak. The signal integral will be referred to as the EDESR amplitude for the remainder of this manuscript. Figure 4 presents the temperature dependence of the EDESR amplitude and line width obtained from the raw data of fig. 3 using the procedure described above. It is apparent that the EDESR amplitude measured in the vicinity of ν=1 has a pronounced maximum at about 2K, decreasing sharply at higher or lower temperatures. As seen in fig. 3 the $\Delta R_{xx}$ non-resonant background component behaves in a similar way, reaching a maximum at the same temperature as the ESR. This behavior is in contrast with that reported in [5] where the disappearance of the ESR with lowering the temperature was accompanied by a growth of the non-resonant background. The line width in fig. 4 appears to be nearly independent of temperature. Assuming that the line width is inversely proportional to the spin-spin relaxation time $T_2$, as in conventional ESR detection by direct microwave absorption, then it may be concluded that $T_2$ is independent of temperature over the temperature range of interest. We have also measured the filling factor dependence of the ESR amplitude (at constant temperature) in the vicinity of ν=1, as shown in fig. 5. The filling factor was varied by incrementally rotating the sample at a fixed microwave power and frequency. Despite the appreciable scatter in the data points, it is apparent that the EDESR signal maximum occurs close to ν=1. Furthermore, a change in sign of the EDESR is observed for filling factors below 0.93 and above 1.08. In general, the EDESR has been found to have the same sign as the slope of $R_{xx}(T)$ at the same filling factor. The EDESR amplitude decreases upon moving away from ν=1 where it is positive and obtains a maximum, passes through zero at the critical points on both sides of the ν=1 minimum where $R_{xx}$ is temperature independent, and then changes sign at both higher and lower filling factors. Finally, fig. 6a presents a typical activated temperature dependence of $R_{xx}$ at $\nu = 1$ in sample EA-124. The activation energy gap determined from this data (see fig. 6b) is 7.0 K, while the single electron spin Zeeman interaction at 5.7 T is equivalent to a 1.6 K gap. The corresponding EDESR amplitude temperature dependence exhibits a maximum near 2.2 K.

**Discussion**

We now propose a mechanism for EDESR whereby excitation of ESR produces a change in the longitudinal magnetoresistance, $\Delta R_{xx}$. In the simple density of states model for a 2D electron system at ν=1 the longitudinal magnetoresistance, $R_{xx}$, passes through a minimum when the chemical potential is located midway between the maxima in the density of states corresponding to the spin-up and spin-down configurations of the lowest Landau level (N=0). The standard method for measuring the spin gap, $\Delta E$, is by an activated transport experiment wherein the temperature dependence of $R_{xx}$ is fit to the Arrhenius law:

$$R_{xx} = R_0 \exp\left(-\frac{\Delta E}{2k_B T}\right) \tag{1}$$

where the prefactor $R_0$ is related to the sample resistance in the limit $\Delta E \ll k_B T$. The gap $\Delta E$ is usually determined from the slope of a plot of $\ln(R_{xx})$ versus $1/T$.

To begin, it should be recognized that the energy gap $\Delta E$ obtained by fitting eq. (1) to the temperature dependence of the longitudinal resistance is not the same energy gap associated with EDESR transitions because, according to Larmor's Theorem, the electron spin Larmor frequency (and the frequency of optical transitions in general) does not depend on Coulomb interactions. In a spin wave system, the microwave field couples only to the $k=0$ magnetic exciton, where according to Kohn's theorem [11], the center-of-mass motion (with $k=0$) is unaffected by electron-electron interactions, so that $\Delta E(k=0) = g\mu_B B_0$. The activated transport experiment, on the other hand, deals with excitations that produce charge carriers via ionization processes. Hence, magnetoresistance probes the large–$k$ limit of the excitation dispersion relation corresponding to a well-separated electron-hole pair. The splittings of the 2D electron system subjected to a perpendicular magnetic field is usually expressed as a sum of the electron spin Zeeman and the electron-electron Coulomb energies:

$$\Delta E = m\hbar\omega_c + \Delta m_s g\mu_B B_0 + \Delta E_c(k, B_0) \tag{2}$$

where $m=1$ for magnetoplasmons and $m=0$, $\Delta m_s = 1$ for spin waves, where $\omega_c$ is the cyclotron frequency and $\Delta E_c(k, B_0)$ is the energy contribution due to the electron-electron interaction [12]. Accordingly, the energy gap probed by thermal activation of transport at ν=1 is $\Delta E = g\mu_B B_0 + \Delta E_c(k \to \infty, B_0)$. For the magnetic fields pertaining to ν=1 in our GaAs/AlGaAs quantum well samples, the Coulomb term in the $k \to \infty$ limit is much greater than the Zeeman term. For example, $|g|\mu_B B_0/k_B = 1.57$ K at 5.7T, while in the ideal 2DES, according to theory, $\Delta E_c(k \to \infty, B_0 = 5.7T)/k_B = 162$ K .[12,15] The experimentally measured gap of $\Delta E/k_B = 7$ K in sample EA-124 is substantially smaller than the theoretical gap. Some of the reasons for reduction in comparison to the theoretical gap include the inclusion of finite $z$ extent, the influence of disorder, and well-to-well variation in electron density in multiple quantum well samples such as EA-124. [12]. Under our experimental conditions, the "spin gap" measured by thermal activation of $R_{xx}$ is much greater than the Zeeman gap corresponding to ESR transitions induced by a spatially uniform (relative to the space scale of the e-e interaction) microwave field. Therefore, a proper description of EDESR at ν=1 in the quantum Hall state must incorporate electron correlation effects. The elementary neutral excitations may be described as quantized *spin waves* (i.e. magnons) or equivalently as *spin excitons* [13,14]. An important feature of these charge neutral excitations is that they occur with a conserved wave vector $\vec{k}$, and the excitation of a spin wave mode corresponds to exactly one electron spin flip which is distributed over many spins.

Our proposal is that electrically detected ESR can be conceptualized as a two step process: (1) the resonant absorption of microwave photons with energy $\hbar\omega_0 = g\mu_B B_0$, and (2) a resultant increase in the internal energy (i.e. heating) of the 2DES which produces $k \to \infty$ excitations that are detected via a change in $R_{xx}$ according to Eq. (1). Our model will incorporate the dispersion relation for a 2D spin wave system at $\nu=1$ which has been derived independently by Byckov [14] and Halperin [15]:

$$\Delta E(k) = g\mu_B B_0 + \frac{e^2}{4\pi\varepsilon_0 \varepsilon l_0}\left(\frac{\pi}{2}\right)^{\frac{1}{2}}\left[1-e^{-k^2 l_0^2/4}I_0\left(\frac{k^2 l_0^2}{4}\right)\right] \quad (3)$$

Here, $l_0 = \sqrt{\hbar/eB_0}$ (SI units) is the magnetic length, $k$ is the magnitude of the wave vector and $I_0$ is the modified Bessel function of the first kind. For computational purposes eq. (3) can be simplified to:

$$\Delta E(x)/k_B = 0.27 B_0 + 68 B^{1/2}\left[1-e^{-x}I_0(x)\right] \quad (4)$$

where $x = k^2 l_0^2/4$. At thermal equilibrium the average number of magnons excited in the mode $k$ is given by the Planck distribution:

$$N_k = \frac{1}{e^{\Delta E(k)/k_B T} - 1} \quad (5)$$

The total number of modes per unit area is equal to the total number of spins (per unit area) contributing to the magnetization in a 2D layer:

$$\sum_k N_k = \frac{1}{(2\pi)^2}\int_0^\infty \frac{1}{\left(e^{\Delta E(k)/k_B T} - 1\right)} d^2k$$

$$= \frac{1}{\pi l_0^2}\int_0^\infty \frac{1}{\left(e^{\Delta E(x)/k_B T} - 1\right)} dx \quad (6)$$

In accordance with Kohn's theorem, only $k=0$ spin wave modes couple to the microwave field, and the spin resonance energy absorption will occur only when the photon energy of the microwave field equals the single electron Zeeman energy, $g\mu_B B_0$. In our model, the temperature change of the 2DES is due to the deposition of the resonant microwave energy absorption into the *k*=0 excitation. Thus, we need to find a relation between the power absorbed by excitation of the uniform mode and the observed change in the longitudinal resistance $R_{xx}$ at ν=1.

At low microwave field, the dynamics of the magnetization can be approximately described by the classical torque due to an effective magnetic field, $\vec{B}_{eff}$ : [16]

$$\frac{d\vec{M}}{dt} = -\gamma\left(\vec{M} \times \vec{B}_{eff}\right) + \text{dissipative term} \qquad (7)$$

where $\gamma = g\mu_B/\hbar$ is the electron gyromagnetic ratio. Energy dissipation can occur either by direct spin relaxation to the lattice or by transfer to the short-wavelength modes followed by relaxation to the lattice. This point will be discussed in more detail below. In the EDESR experiment a linearly polarized microwave field with amplitude $b_1$ is applied to the sample. This field can be decomposed into a left and right circularly polarized component, where the component resonant with the precessing magnetization has the form $\vec{B}_1 = \left(\frac{1}{2}b_1\cos\omega t, \frac{1}{2}b_1\sin\omega t, 0\right)$. The steady state transverse, complex magnetization, given by

$$M_0^+ = \frac{\gamma M_z b_1/2}{\omega_0 - \omega + i\gamma\Delta B_{1/2}/2}, \qquad (8)$$

is related to the susceptibility by $\chi = \chi' - i\chi'' = M_0^+/(b_1/2)$, where $\chi''$ represents the loss by the sample and $\Delta B_{1/2}$ is the full-width of the resonance curve at half maximum. Thus,

$$\chi''(\omega) = \frac{\gamma^2 M_z \Delta B_{1/2}/2}{(\omega-\omega_0)^2 + \gamma^2(\Delta B_{1/2}/2)^2} \qquad (9)$$

At sufficiently low microwave power, we may make the approximation $M_z \approx M_z^{eq}$ (for example, the low-power criterion is $(\gamma b_1)^2 T_1 T_2 \ll 1$ assuming a Bloch-Bloembergen dissipation term in eq. (7)). The steady state microwave power absorbed per unit area of the 2D electron spin system may be calculated from

$$\bar{p} = \frac{\overline{dQ}}{dt} = \frac{1}{4}\chi''\omega b_1^2 \qquad (10)$$

On resonance, this simplifies to $\bar{p} = M_z^{eq} b_1^2 \omega_0/(2\Delta B_{1/2})$.

The magnetization $M_z^{eq}$ can be expressed in terms of the thermal equilibrium electron spin polarization, $P_z(T) = (N_\uparrow - N_\downarrow)/(N_\uparrow + N_\downarrow)$ where $N_\uparrow$ and $N_\downarrow$ are the numbers of spin-up and spin-down electrons. Hence, the total magnetization per unit area is $M_z^{eq}(T) = \frac{1}{2} g\mu_B N_s P_z(T)$. Two different models for the two-dimensional electron system at ν=1 will be considered. In the first, the thermal equilibrium spin polarization of the spin wave system is calculated from [17]

$$P_z(T) = \left[1 - \frac{2}{N_s} \sum_k N_k\right] \tag{11}$$

where $N_s$ is the electron spin density and $\sum_k N_k$ is given by eq. (6). For comparison, the second model to be considered is the spin polarization of a non-interacting (paramagnetic) electron spin system,

$$P'_z(T) = \tanh\left(\frac{g\mu_B B_0}{4k_B T}\right) \tag{12}$$

Equations (11) and (12) are plotted as a function of temperature in fig. (7).

The on-resonance microwave power absorbed by the 2DES is $\bar{p} = \hbar B N_s P_z \gamma^2 b_1^2 / 4\Delta B_{1/2}$ and can now be calculated as a function of temperature. From the EDESR spectrum obtained for EA-124 at 5.7 T and 1.5 K yields $\Delta B_{1/2} = 20 \pm 3 mT$. As is evident from fig. (4) the resonance line width is nearly temperature independent over the temperature range studied. Although in our experimental arrangement there is no way to calibrate the transverse microwave field at the sample, as a rough estimate we take the upper limit to be $b_1 = 1mT$. The power absorbed from the microwave field by the spin system at spin resonance at 5.7 T and 1.5 K: in the spin wave model, $\bar{p} = 6.6 \, mW/m^2$; while in the non-interacting electron model, $\bar{p} = 1.7 \, mW/m^2$.

Now that we have an expression for the microwave power absorbed, a heat equation can be written:

$$\frac{dU_s}{dt} = C_s \frac{dT_s}{dt} = \bar{p} - K_b(T_s - T_b) \tag{13}$$

As indicated in fig. 8, $T_b$ is the temperature of the bath, $T_s$ is the temperature of the spin wave system, and $K_b$ is the thermal conductance that determines the rate of heat flow to the surroundings. The steady state temperature can be obtained by setting $dU_s/dt = 0$, yielding

$$\Delta T = T_s - T_b = \bar{p}/K_b. \tag{14}$$

This establishes the relationship between the change in temperature of the spin wave system and the microwave power dissipated into the spin system by magnetic resonance absorption.

According to Eq. (16), the variation $\Delta R_{xx}$ must be the result of a change in the temperature of the current carriers, which are the magnetic excitations with $k \to \infty$. For an infinitesimal temperature change, $\delta T$,

$$\delta R_{xx} = R_0 \frac{\Delta E}{2kT_b^2} \exp\left(-\frac{\Delta E}{2kT_b^2}\right) \delta T \tag{15}$$

or more generally,

$$\Delta R_{xx} = R_0 \left[ \exp(-\Delta E / 2T_s) - \exp(-\Delta E / 2T_b) \right] \tag{16}$$

To continue, we make two assumptions regarding the manner in which the magnetic resonance response is observed. The EDESR spectrum is acquired by CW microwave excitation at a fixed frequency while the external magnetic field is swept through the resonance. Thus, the "slow passage" condition is obtained whereby the system has sufficient time to come to a steady state spin polarization. That this condition is obtained is supported by the observation that both the shape and the amplitude of the ESR peak are insensitive to a slight variation in the magnetic field sweeping rate. Secondly, it is assumed that the electron spin system (all spin wave modes) is describable by a single temperature, $T_s$. This assumption is difficult to verify because even in the steady state it still might not be possible to ascribe a single temperature $T_s$ to all spin excitations. The distribution of energy among the spin wave modes reached in the steady state will result from a competition between the rate of energy transfer between spin waves and the spin wave system-to-thermal bath energy transfer rates. To analyze this situation in detail is a difficult theoretical problem that has yet to be addressed for a 2DES at $\nu = 1$. Nevertheless, it is possible to proceed if we make the simplifying assumption that a steady state is obtained under ESR excitation that *can* be described by a single temperature $T_s$.

The idea of heating of the spin wave system through the resonant microwave absorption and redistribution to the $k \neq 0$ modes by various scattering mechanisms [18] leads naturally to a heating mechanism for $R_{xx}$ detection of ESR. This model predicts an increase in $R_{xx}$ due to an increase in the spin temperature according to eq.(15), a prediction that is in agreement with our experimental data. Furthermore, the model also can be used to calculate the temperature dependence of the EDESR response. One can see immediately from eq.(15) that the slope $dR_{xx}/dT_b$ is maximized at a bath temperature $k_B T_b = \Delta E/4$, and vanishes as $T_b \to 0$ or $T_b \to \infty$. A similar behavior is observed in the temperature dependence of the ESR response, as shown in fig.(9). The experimental ESR amplitude has a maximum at about 2.2 K in a sample state where the $\nu = 1$ activation gap is $\Delta E/k_B = 7K$. Therefore, a heating model is consistent with the experimental data.

To calculate the steady state temperature increase of the 2DES from eq. (14), a value for the thermal conductance to the surrounding thermal bath is needed. Rather than attempting to calculate $K_b$, we will use a value that is self-consistent with the experimentally observed increase in the resistance at 1.5 K and 5.7 T according to eq. (1). The increase in $R_{xx}$, corresponds to a temperature change $\Delta T \approx 10 mK$. This steady state temperature change, together with the estimate of the power dissipated, allows an estimation of the thermal conductance. For example, $K_{bath} = \bar{p} / \Delta T_{esr} \approx 0.66 \text{ W K}^{-1}\text{m}^{-2}$ for the spin wave model. Assuming that $K_b$ remains roughly constant over the temperature range 0.3-4.2 K and using eq. (14), the temperature dependence of $\Delta T_{esr}$ can now be calculated. The calculated temperature dependence of the EDESR amplitude is plotted in fig. 9 for both the spin wave and the non-interacting models. The model based on spin waves assumes the theoretical result [12,15] of eq. (4) where $\Delta E_c (k \to \infty) = 68 B^{1/2}$. The qualitative agreement of either model with the experimental data confirms the validity of the heating mechanism for EDESR. This is in contrast to the conclusion reported previously [5], where data showing the decrease in the EDESR amplitude with decreasing temperature was interpreted as a depolarization of the $v = 1$ state at low temperatures due to the correlation effects. Our theoretical model indicates that the position of the maximum ESR response depends primarily on the activation energy gap, $\Delta E$, determined from transport, but as is also evident from fig. (9), electron correlation do affect the position of the maximum and the shape of the temperature dependence.

**Conclusions**

The electrically detected electron spin resonance amplitude and line width have been measured as a function of both temperature for $T$=0.3-4.2 K and filling factor in the vicinity of $v = 1$ in a 2D electron system in GaAs quantum wells. The EDESR signal is observed as a sharp peak in $\Delta R_{xx}$ when the photon energy of the microwave field is resonant with the Zeeman energy splitting associated with the bare $g$ factor of the electron. The EDESR line width is nearly constant in the temperature range studied, while the EDESR amplitude has a maximum at $k_B T \approx \Delta E / 4$, where $\Delta E$ is the exchange enhanced spin gap determined from thermal activation of transport at $v = 1$. While the position of the maximum in the EDESR amplitude is sensitive to the nature of the excitations at $v = 1$, the occurrence of a maximum and the disappearance of the signal as T→0 is predicted by a heating mechanism in either the independent electron or spin wave models for the 2DES. This is in contrast to the conclusion made in an earlier report wherein the disappearance of the EDESR signal at low temperature was explained in terms of spin depolarization of the $\nu = 1$ state. The proposed heating model correctly predicts the location of the maximum in the experimentally observed temperature dependence of the EDESR amplitude. It also correctly predicts that the signal should vanish as the temperature is increased or decreased. The results of the present study demonstrate how experimental EDESR studies can, under appropriate conditions, provide data that can be used to discriminate between competing theories for the magnetic ordering and magnetic excitations of a 2DES in the regime of the quantum Hall effect.

**Figure captions**

Fig. 1. Block diagram of the experimental setup for EDESR measurements on a GaAs quantum well Hall bar using the double lockin technique described in the text.

Fig. 2. (a) Typical $R_{xx}$ vs. $B_0$ trace for sample EA-124 at $T$=1.5 K. The $\nu=1$ minimum occurs at about 5.7 T. (b) Typical $\Delta R_{xx}$ vs. $B_0$ response at a temperature of $T=1.5K$. The sample is tilted at $60°$ with respect to the external magnetic field. The EDESR signal appears as a sharp peak at $B_0 = 5.47$ T. The inset shows the ESR feature on an expanded scale. The microwave frequency is 32 GHz, which at this magnetic field corresponds to $|g| = 0.418$.

Fig.3. The experimental $\Delta R_{xx}$ vs. $B_0$ curves show at several selected temperatures, as indicated. The non-resonant background contribution to $\Delta R_{xx}$ has not been removed from this raw data. Both the ESR and the non-resonant background contributions to the signal exhibit maximum amplitudes at about 2.2 K.

Fig. 4. Temperature dependence of (a) the ESR peak amplitude and (b) the ESR line width. The amplitude and line widths were extracted from the data shown in fig. 2 after removal of the non-resonant contribution to the $\Delta R_{xx}$ vs. $B_0$ signal according to the procedure described in the text. Data were recorded for magnetic field up sweep and magnetic field down sweep to confirm that the effect due to dynamic nuclear polarization and Overhauser shift is negligible at the field sweep rate of 0.448 Tesla/min used to record the EDESR spectra.

Fig.5. (a) The ESR amplitude as a function of filling factor in the vicinity of ν=1 at $T$=1.5 K. The ESR has a maximum at roughly ν=1 and changes sign for ν<0.94 and ν>1.075. The sign of the ESR is correlated with the sign of the temperature dependence of $R_{xx}$, supporting the proposal that the EDESR mechanism involves a heating effect. (b) Calibration of the filling factor for part (a). The curves represent segments of $R_{xx}(B_0)$ in the vicinity of ν=1 for each tilt angle. To obtain the ESR signal as a function of filling factor the sample was incrementally rotated for each ESR measurement. The microwave frequency was chosen to produce an ESR peak at 5.65 T, as indicated by the dashed vertical line.

Fig.6. (a) Temperature dependence of $R_{xx}(B_0)$ in the vicinity of ν=1 at $B_0$=5.7 T. (b) Arrhenius plot of $\ln(R_{xx})$ versus $1/T$ at the ν=1 minimum which yields a spin splitting of $\Delta E/k_B = 7K$.

Fig. 7. Theoretical temperature dependence of the spin polarization, $P_z(T)$, at ν=1 and $B_0$=5.7 T for a 2DES in GaAs consisting of either non-interacting electrons (eq. (12)) or spin waves (eq. (11)).

Fig.8. Electrical circuit diagram equivalent showing energy flow from the microwave field to the spin wave system and thermal bath. As explained in the text, $\bar{p}$ is the power absorbed by the $k$=0 excitations, while $K_b$ is the thermal conductance from the sample at a steady state temperature $T_s$ to the thermal bath at temperature $T_b$.

Fig.9. Experimental data showing the temperature dependence of the EDESR amplitude (after removing the non-resonant contribution to $\Delta R_{xx}$) at 5.7 T. Data points represented by the filled circles correspond to spectra recorded with a magnetic field downsweep while open circles represent data recorded with magnetic field up-sweeps. The $\nu=1$ state is characterized by an activation energy gap of $\Delta E/k_b = 7$ K, as determined from the Arrhenius plot in fig.6. The continuous curves represent calculations for a 2DES assuming either non-interacting electrons (solid curve, see eq. (12)) or a spin wave model (dashed curve, see eq. (11)).

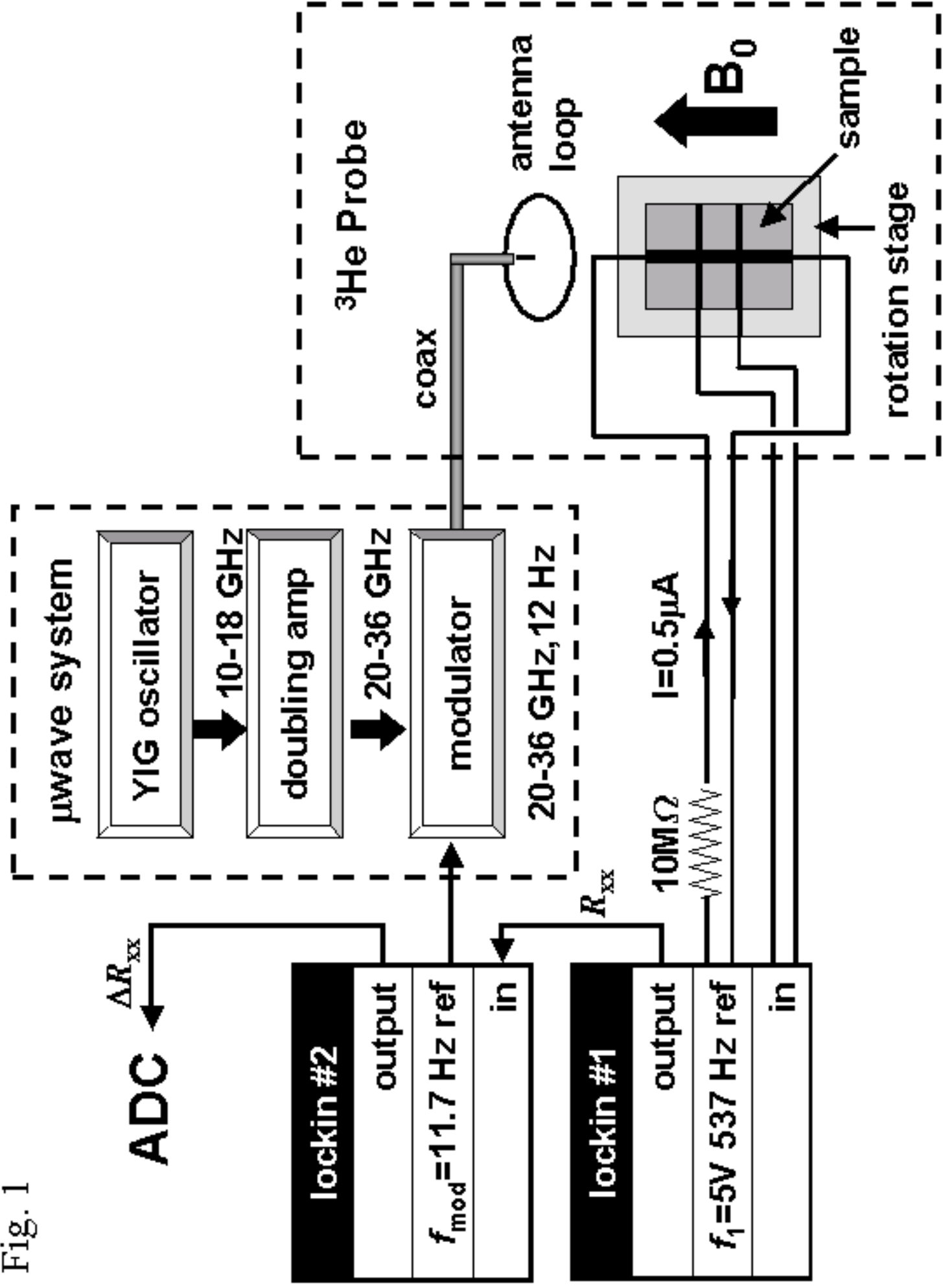

Fig. 1

Fig. 2.

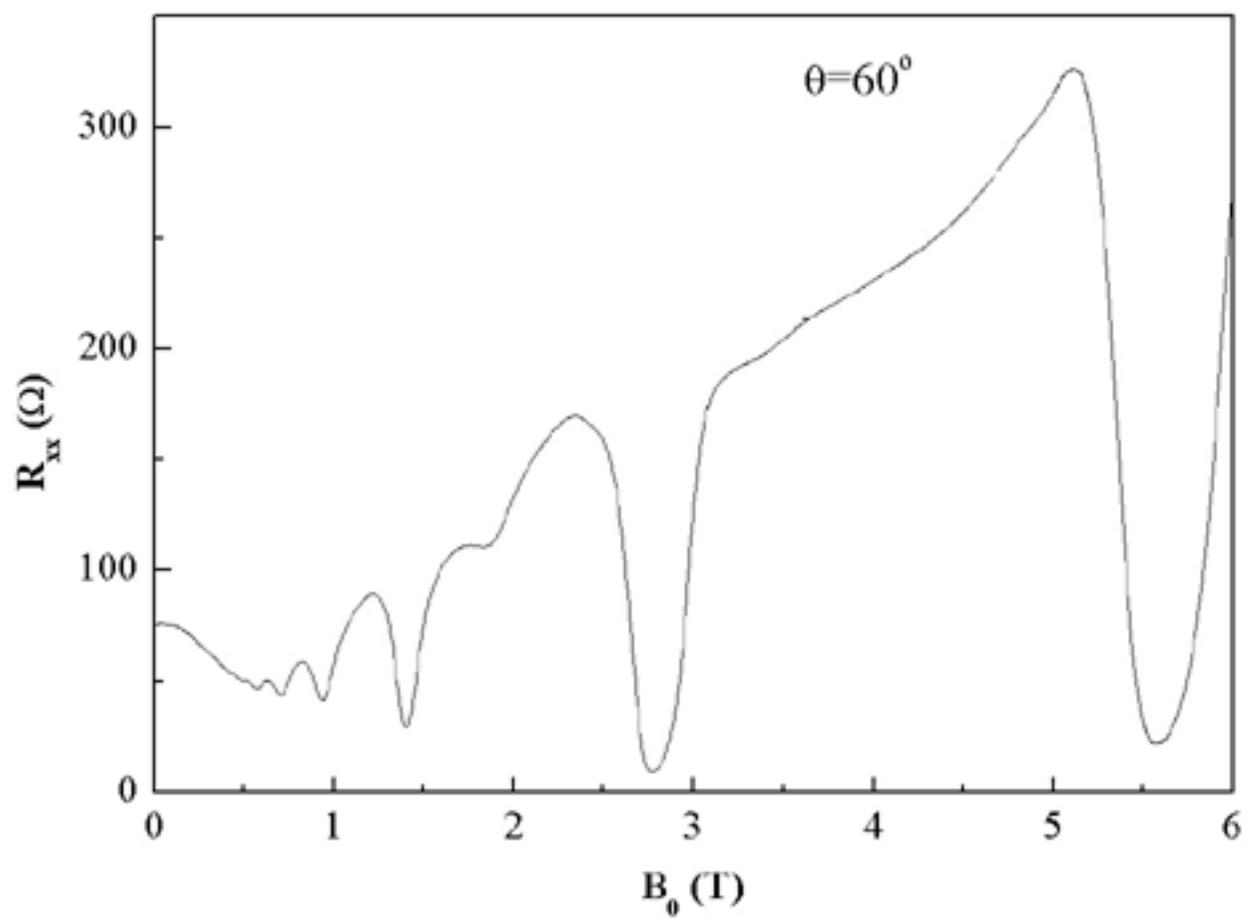

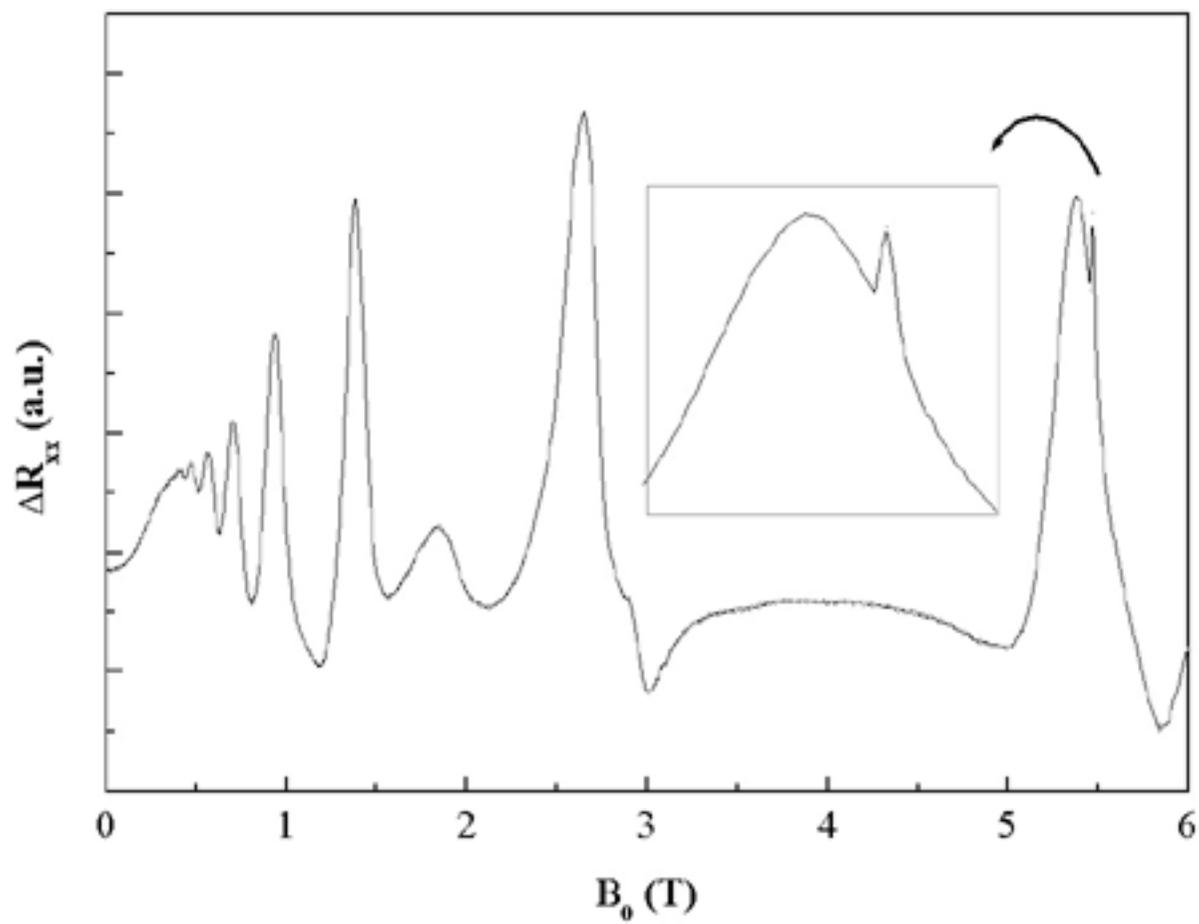

Fig. 3.

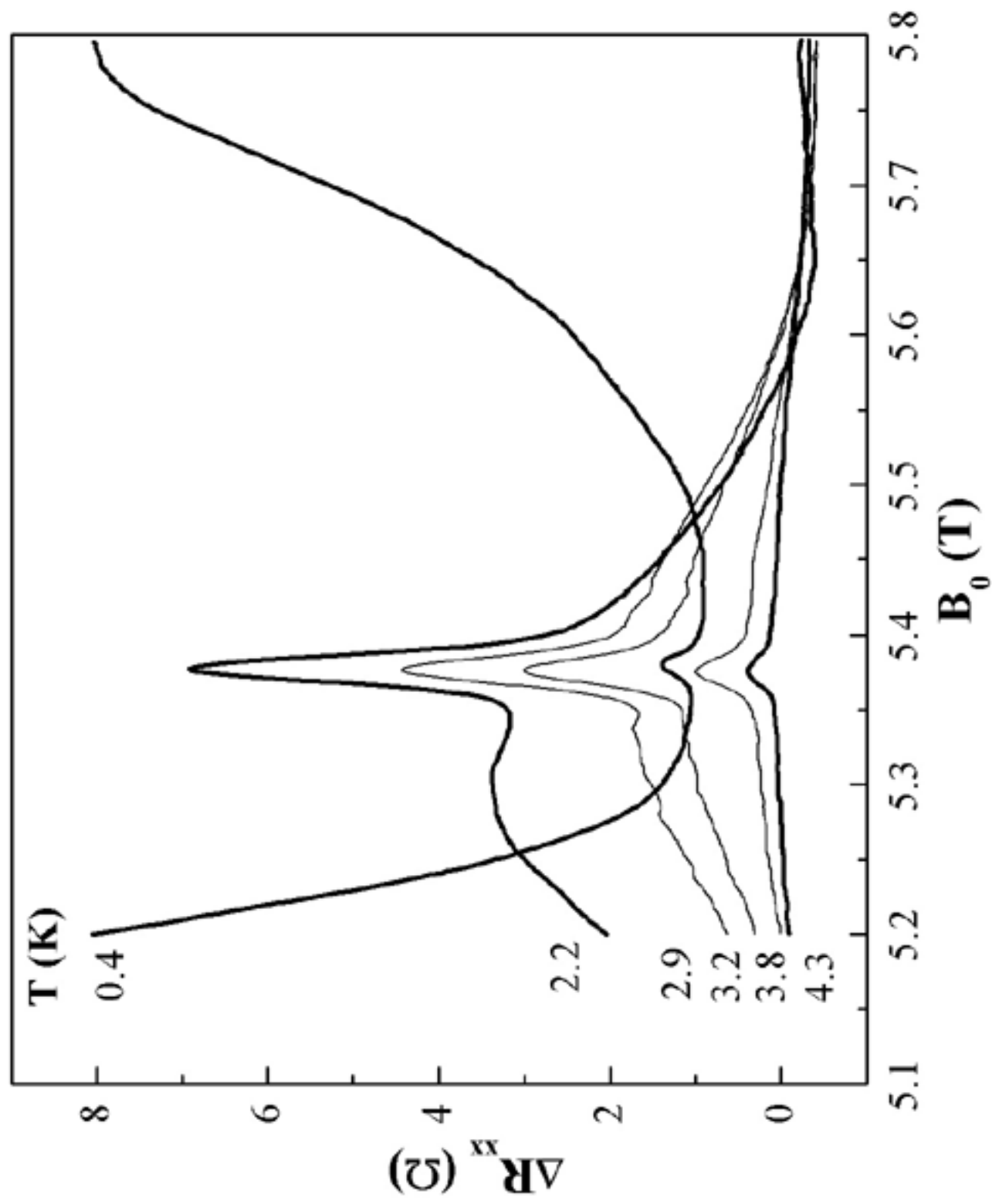

Fig. 4.

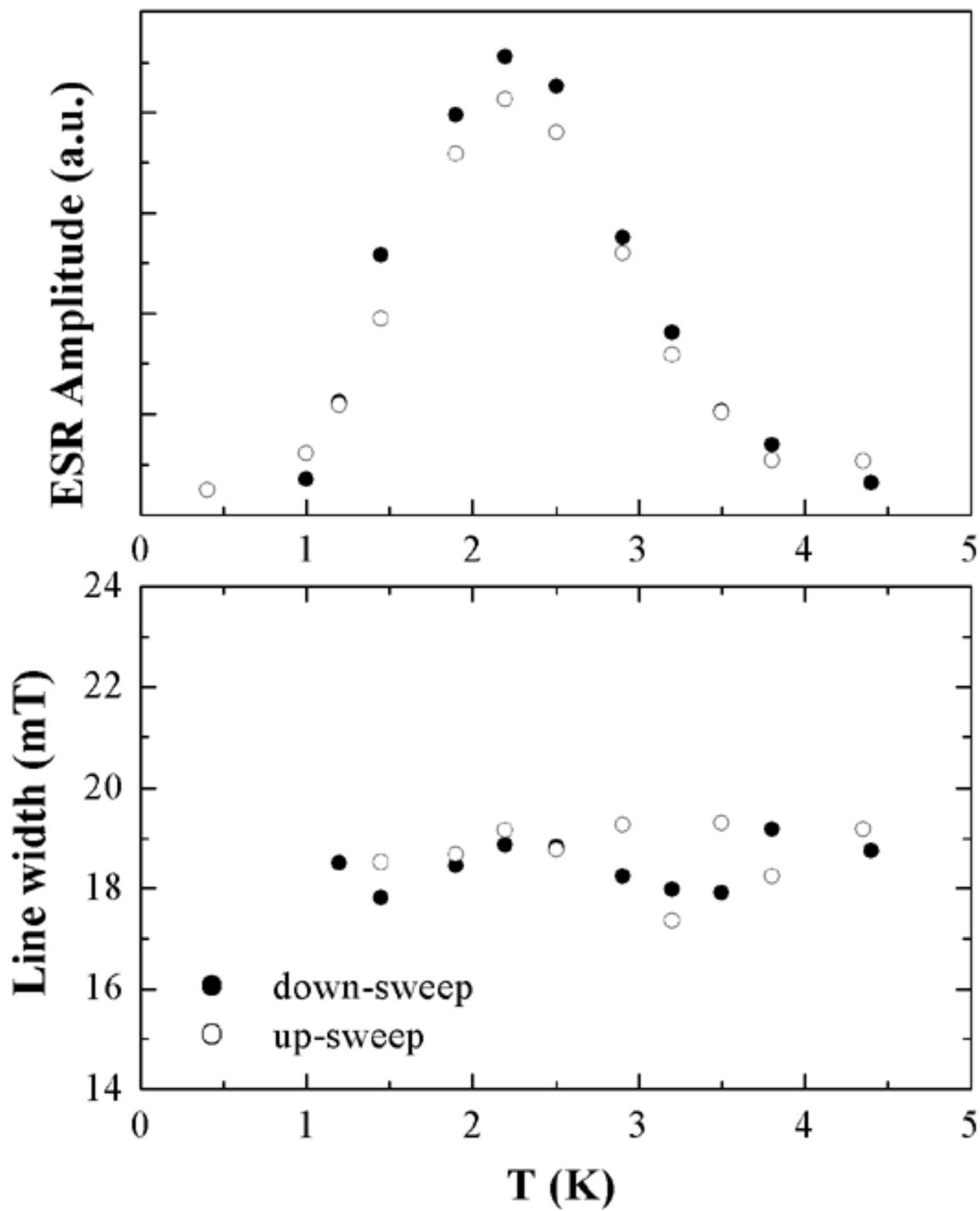

Fig. 5.

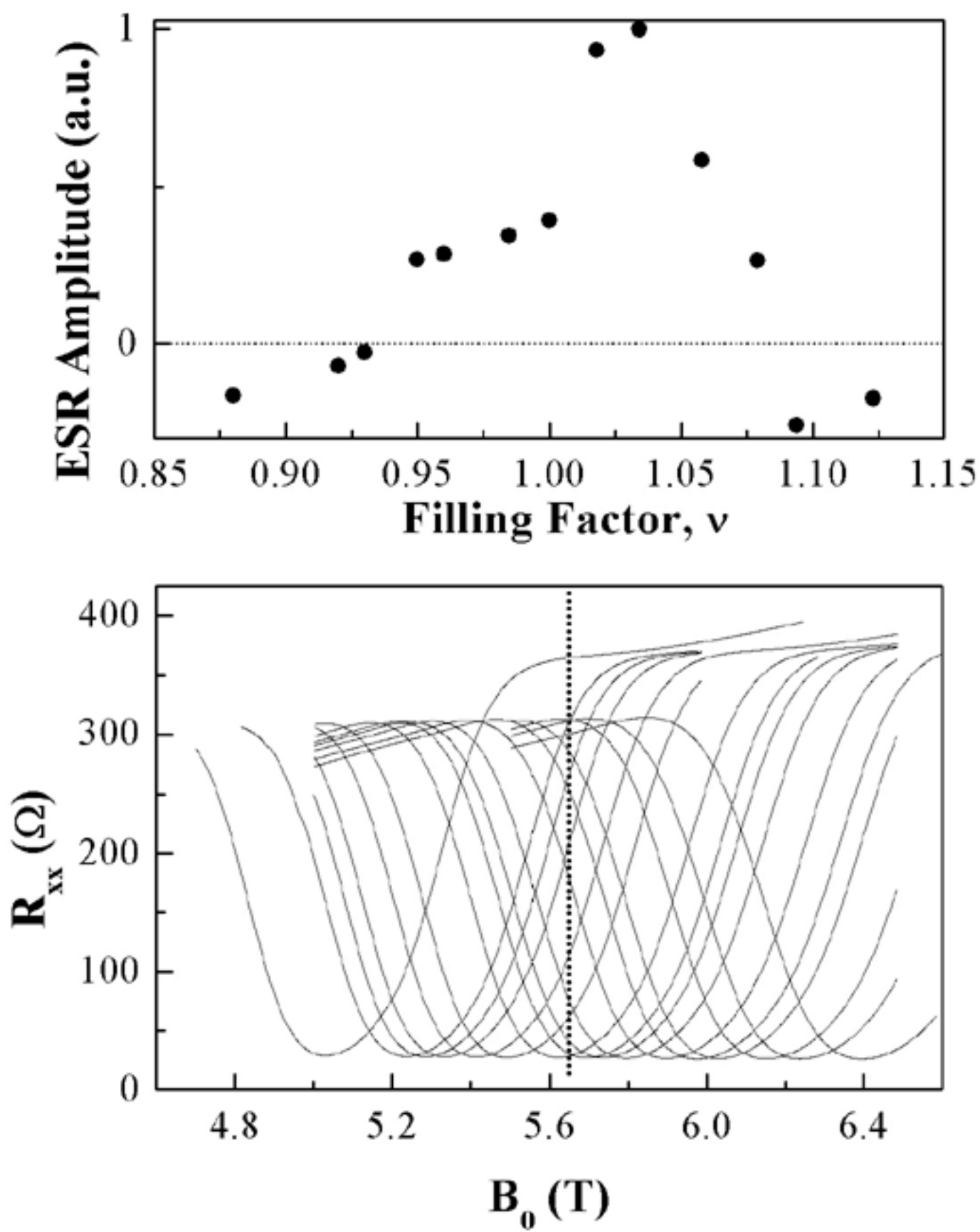

Fig. 6.

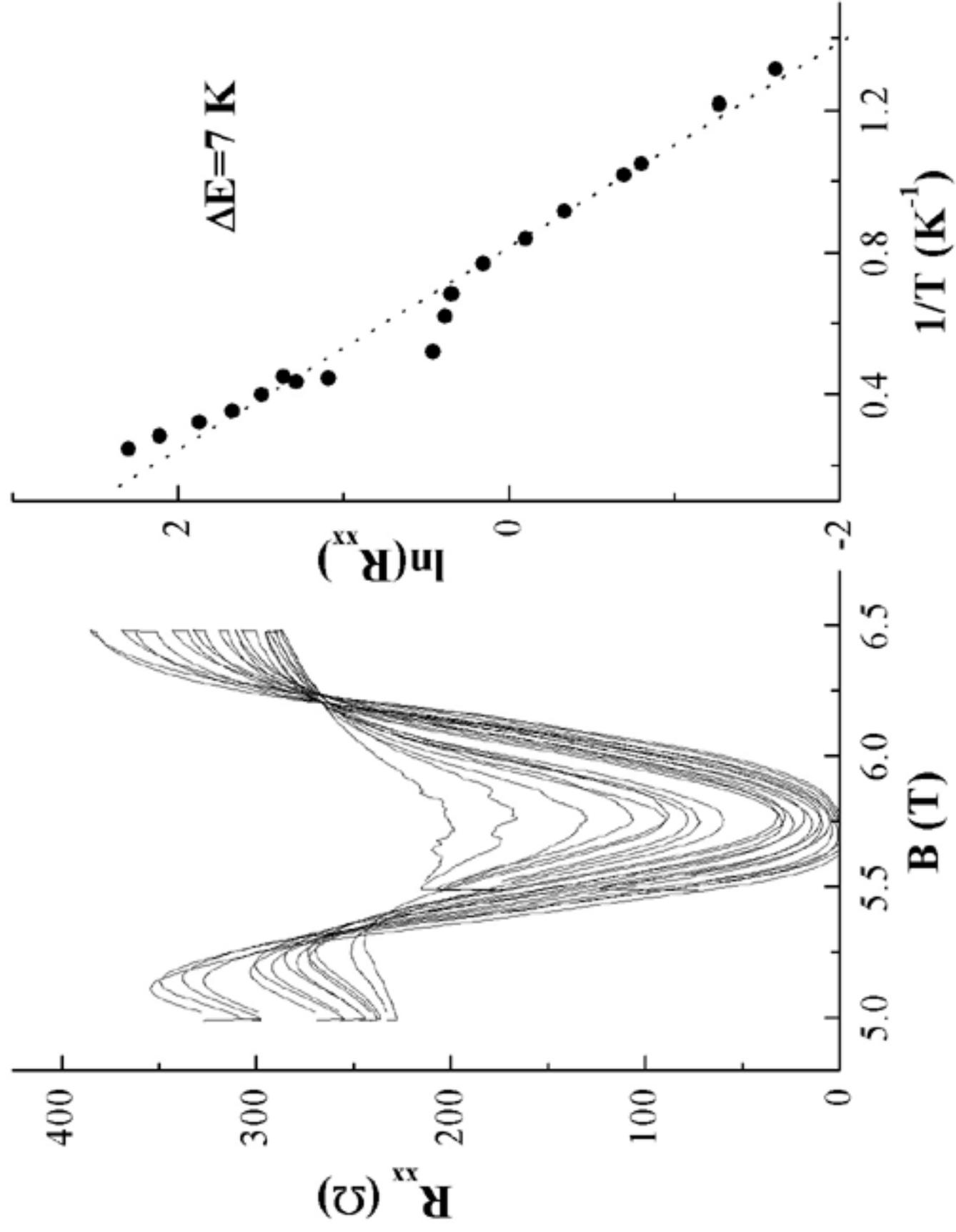

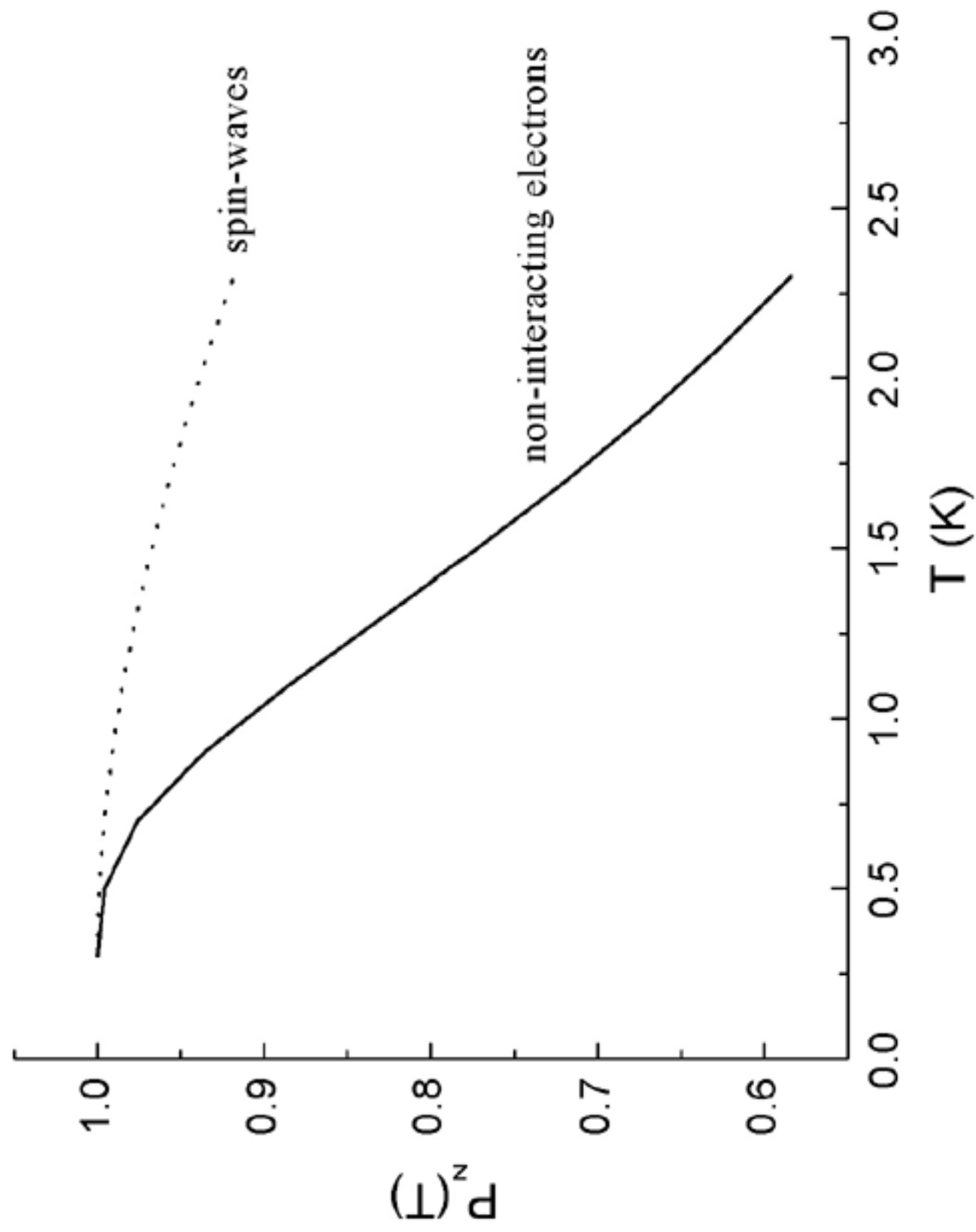

Fig. 7.

Fig. 8

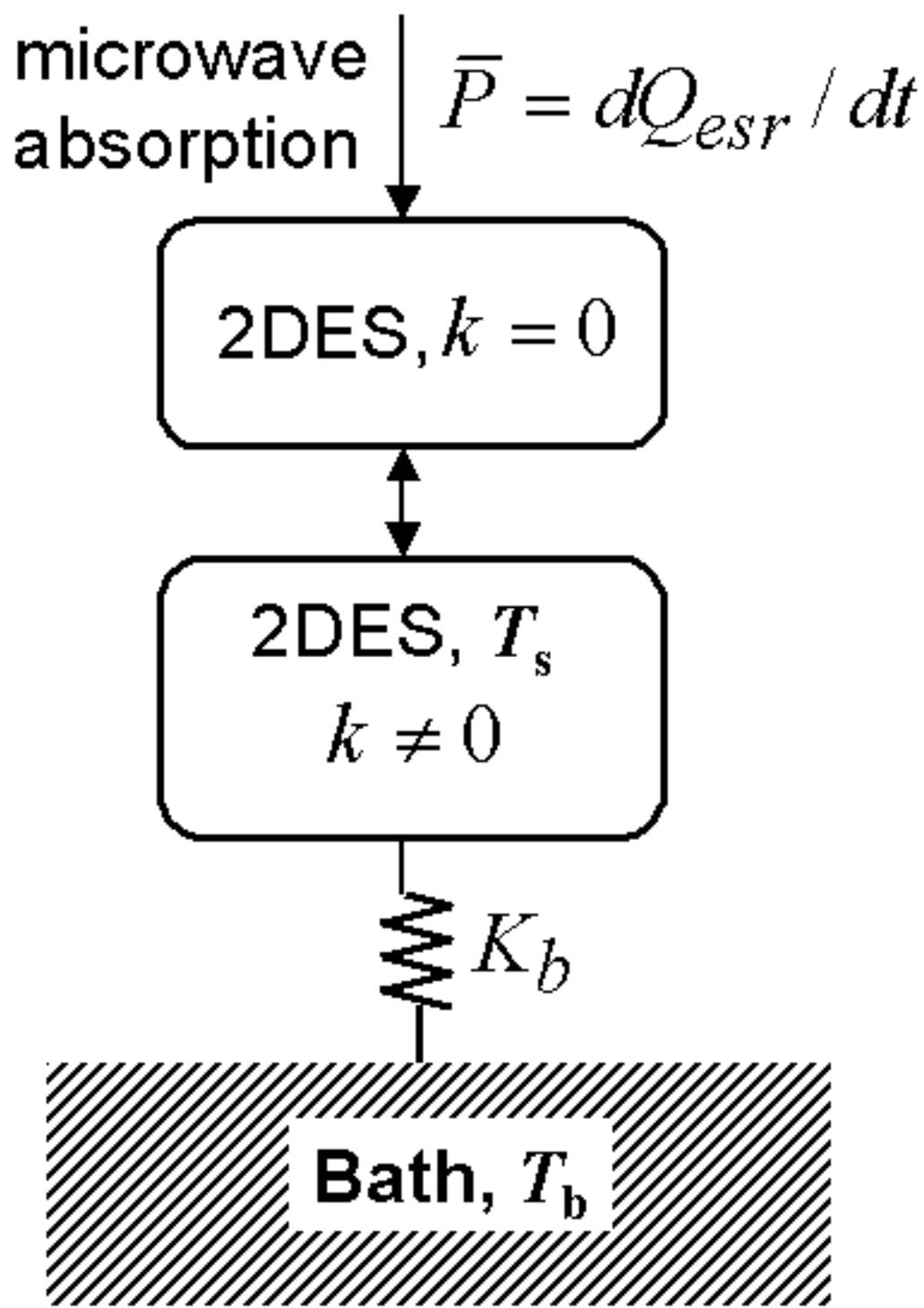

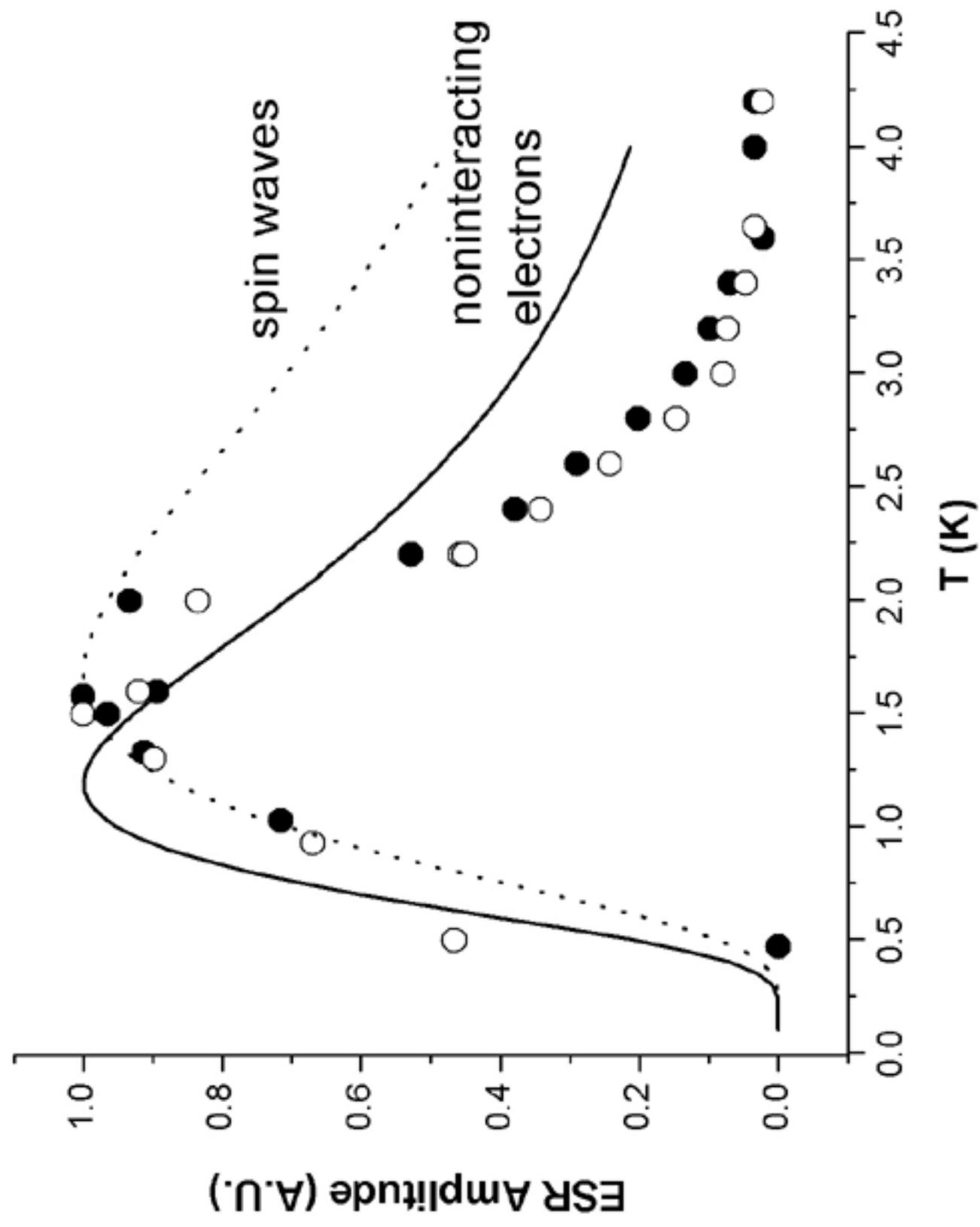

Fig. 9